\documentclass[prl,twocolumn,floatix,amssymb,showpacs,amsmath,superscriptaddress,usenames,dvipsnames]{revtex4-1}

\usepackage[T1]{fontenc}
\usepackage[svgnames]{xcolor}
\usepackage[colorlinks=true,
            linkcolor=DarkRed,
            urlcolor=Magenta,
            citecolor=Blue,
            anchorcolor=red]{hyperref}
\usepackage{amsmath, dsfont, bm}
\usepackage{physics,braket}
\usepackage{graphicx}
\usepackage{mathptmx}
\usepackage{color}
\usepackage[normalem]{ulem}
\usepackage[makeroom]{cancel}

\begin{document}

\title{Enhancing Gravitational Interaction between Quantum Systems by a Massive Mediator}

\author{Julen S. Pedernales, Kirill Streltsov, and Martin B. Plenio}
\affiliation{Institut f\"ur Theoretische Physik und IQST, Albert-Einstein-Allee 11, Universit\"at Ulm, D-89081 Ulm, Germany}

\begin{abstract}
In 1957 Feynman suggested that the quantum/classical character of gravity may be assessed by testing the gravitational interaction due to source masses in superposition. However, in all proposed experimental realisations using matter-wave interferometry the extreme weakness of this interaction requires pure initial states with extreme squeezing to achieve measurable effects of non-classical interaction for reasonable experiment durations. In practice, the systems that can be prepared in such nonclassical states are limited to small masses, which in turn limits the strength of their interaction. Here we address this key challenge---the weakness of gravitational interaction---by using a massive body as an amplifying mediator of gravitational interaction between two test-systems. Our analysis shows that this results in an effective interaction between the two test-systems that grows with the mass of the mediator, is independent of its initial state and, therefore, its temperature. This greatly reduces the requirement on the mass and degree of delocalization of the test systems and, while still highly challenging, brings experiments on gravitational source masses a step closer to reality.
\end{abstract}

\maketitle

{\em Introduction.---}In a discussion regarding the necessity of gravitational quantization at the 1957 Chapel Hill 
Conference on the Role of Gravitation in Physics Richard Feynman, aiming  to clarify a point made by Frederik 
Belinfante, presented a Gedanken experiment in which a coherent superposition of a massive particle in two different spatial locations, generated e.g., by a particle in a coherent superposition of spin states 
entering a Stern-Gerlach apparatus, is allowed to interact gravitationally with another mass~\cite{Feynman57}. He pointed out that the two possibilities for treating the gravitational interaction, either via a classical or via a quantum field, result in very different quantum 
states and thus experimental outcomes. Notably, the particles would, respectively, emerge in a classically 
correlated mixture of different positions or in a coherent superposition. The latter case is, in modern quantum information parlance, referred to as an entangled state.

At the time such a Gedanken experiment was extraordinarily far removed from the experimental technology of the day. 
After all, it was only in 1952 that Schr{\"o}dinger wrote ``... we never experiment with just one electron or 
atom or (small) molecule. In thought experiments, we sometimes assume that we do; this invariably entails ridiculous 
consequences [...] we are not experimenting with single particles, any more than we can raise Ichthyosauria in the 
zoo''~\cite{Schrodinger52}. Owing to this evident technological gap, there has been little activity by experiment and theory to explore possible 
routes towards turning Feynman's Gedanken experiment into reality.

However, six decades later, the rise of advanced quantum technologies and, notably, the field of optomechanics is starting to change this perception. The increasing ability to bring particles of ever growing mass into 
the quantum regime and control their dynamics in a manner that leaves their coherences intact~\cite{AspelmeyerKM14,TeufelDL2011, ChanMS2011, RiedingerWM2018, OckeloenDP2018, MillenMP2020, DelicRD2020, CatanoSE2020, Whittle21, KotlerJT+2021, MercierMS+2021} suggests that today such an experiment may be conceivable albeit still extraordinarily challenging \cite{SchmoleDH+16}. Indeed, by determining experimentally the entanglement gain between two gravitationally interacting parties one would be able to falsify the assumption of a classical force carrier and thereby conclude the non-classical nature of the gravitational field between them~\cite{KafriT13,KrisnandaTP2017}. This led to further proposals for experiments that probe for gravitationally induced entanglement~\cite{Milburn2017,MarlettoVV17,MiaoMY+20,KrisnandaTP+20,PedernalesMP20b,PedernalesMP20,CoscoMP20,Weiss2020} and add to other tests based on superpositions of source masses~\cite{LindnerP2005,BahramiBM+15,CarlessoPU+17,CarlessoHU19,Haine2021}. While these experiments might become feasible at some point, it is equally clear that remarkably stringent requirements on isolation from the environment, the required duration of these experiments and the large spatial extent of the quantum superpositions that are required to achieve a measurable effect render such type of experiment extremely challenging indeed~\cite{PedernalesMP20}.
\begin{figure}[htbp] %  figure placement: here, top, bottom, or page
   \centering
   \includegraphics[width=\columnwidth]{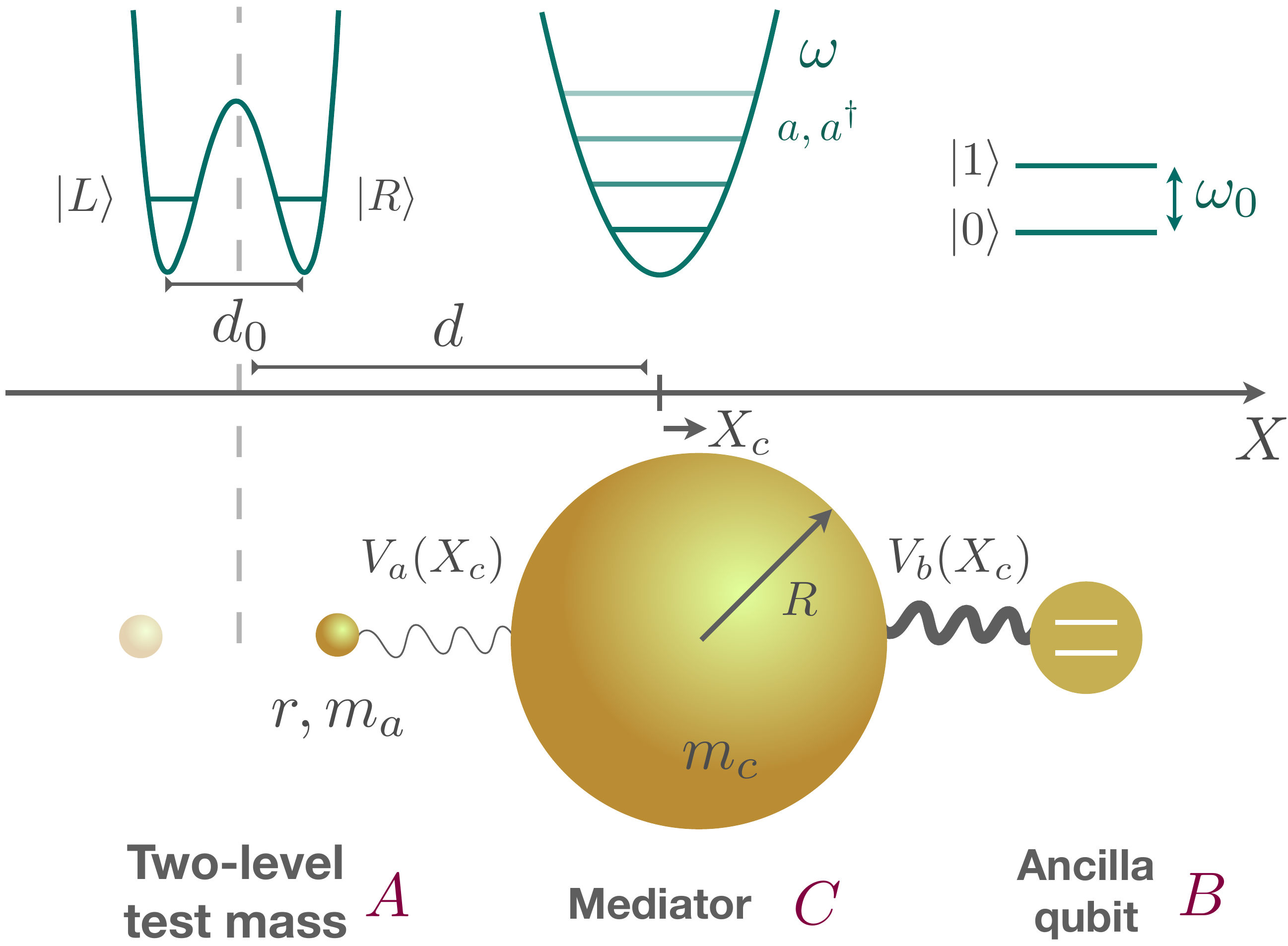} 
   \caption{{\bf Setup} A test particle (system $A$) of radius $r$ and mass $m_a$ is subject to a double-well potential with wells separated by a distance $d_0$ and behaves as a two-level system with states $\ket{L}$ and $\ket{R}$, which are stationary for the duration of the experiment provided that $d_0$ is large enough to make any tunnelling negligible. A massive oscillator (system $C$) with frequency $\omega$, radius $R$ and mass $m_c$ has its equilibrium position at a distance $d$ from the center of the double-well potential and acts as a mediator between the test mass and an ancillary qubit (system $B$) that has states $\ket{0}$ and $\ket{1}$ and bare energy splitting $\omega_0$. The mediator is weakly coupled to the test mass through gravitational interaction with energy $\hat{V}_a(\hat{X})$ depending on the position of the oscillator, and strongly coupled to the ancillary system with a much stronger interaction energy $\hat{V}_b(\hat{X})$ of a nature other than gravitational, e.g., Casimir force. The direct interaction between systems $A$ and $B$ is negligible.  }
   \label{fig:setup}
\end{figure}

In this Letter, we show that by introducing a heavier mediator particle that interacts gravitationally with a smaller test mass and by some other stronger force with an ancillary quantum system, an effective interaction between the test mass and the ancillary system can be engineered which grows with the mass and degree of delocalization of the mediator. Notably, at suitably chosen points in time, the mediator decorrelates from the system, leaving only the test mass and the ancillary system entangled. As a result we find that, a light test mass can be made to interact with an ancillary system as if it had the much larger mass of the mediator, with the significant benefit that the heavier mass of the mediator need not be prepared in a pure state and can, thus, remain at a finite temperature. Since a key technological challenge resides in the difficulty of preparing a sufficiently heavy mass in a pure state with a large enough spatial extension, the setup described here represents a significant enhancement over existing proposals.

{\em Concept and Setup.---}For the calculations in this paper, we will assume a gravitational interaction strength that is determined by the Newtonian interaction energy, which for two bodies of mass $m$ with their centers of mass (c.m.\@) located at positions $x_1$ and $x_2$ is given by $E_G = -G m^2/\abs{x_1 - x_2}$. Expanding this for small variations of $\abs{x_1 - x_2}$ around a fixed separation distance $d$, we find that the lowest-order coupling term is linear in the positions of the two masses and has the form $Gm^2 x_1 x_2/d^3$. Under such an interaction the c.m. of the two particles entangle at a rate that grows with the extent of their spatial delocalization~\cite{KafriT13,Milburn2017, CarlessoPU+17,KrisnandaTP+20,PedernalesMP20,CoscoMP20,Weiss2020}.
While this describes gravity as a direct interaction, ignoring any field degree of freedom that may mediate the force, it allows for the computation of the attainable amount of entanglement, the presence of which may then allow us to draw inferences regarding the classical or quantum character of gravitational interaction and the field that may be mediating it.

For two masses that are trapped in local harmonic potentials and cooled down to their motional ground states (GS), the amount of entanglement due to their gravitational interaction, as quantified by the logarithmic negativity~\cite{Plenio05}, oscillates in time with its maximum given by $\eta = 2Gm/(\omega^2 d^3)$ at time $t = \pi / [(1-\eta)2\omega]$  \cite{KrisnandaTP+20}. To ensure that the gravitational interaction dominates over Casimir forces, the surface-to-surface distance between the interacting bodies must be kept above a certain threshold determined by the radii of the particles. Interestingly, for large particles, when the separation distance is dominated by their size, $\eta$ becomes independent of the particle size. This appears to be a strong limitation, as the gravitational interaction is naturally minute, and it seems the amount of entanglement that it can generate cannot be enhanced above a certain threshold even if we would acquire the ability to cool down objects of larger size~\cite{StreltsovMP21}. 
One way to avoid this limitation is to increase the spatial extent of the c.m.\@ wavefunctions above that of their GSs, for example by squeezing them \cite{KrisnandaTP+20,PedernalesMP20,Weiss2020,CoscoMP20} or by placing each system in a superposition of two spatially separated coherent states~\cite{Milburn2017,PedernalesMP20}. However, the entanglement generated will be extremely sensitive to the tiniest decoherence sources of the involved systems~\cite{PedernalesMP20,vandenKamp2020,Torov2020,RijavecCM21} and, in general, this sensitivity will grow with increasing delocalization of the system~\cite{PedernalesMP20,CoscoMP20}. Therefore, the challenge for the observation of gravitationally induced entanglement resides in the ability to generate highly nonlocalized states of massive objects whose purity needs to be maintained over the duration of the protocol. This is a phenomenal technological challenge that increases with the size of the objects. In the remainder of this paper, we introduce and analyse a setup where the requirement of having a heavy mass in a highly delocalized state is not imposed on the test masses that we want to entangle but is instead shifted onto a third system that serves to mediate their interaction. While the test systems require their preparation in suitable pure states, the mediator can take any pure or mixed state, and the effective interaction between the test systems can be enhanced by increasing the size of the mediator instead of that of the test systems themselves. 

Consider the setup depicted in Fig.~\ref{fig:setup} consisting of three interacting systems, $A$, $B$ and $C$. We denote system $A$ as a two-level test mass (TLTM), i.e. it is a particle of mass $m_a$ trapped in a double-well potential along dimension $X$ and behaves as a two-level system with states $\ket{L}$ and $\ket{R}$, which correspond to the particle being located, respectively, in the left or in the right well. We assume that the wells are deep and far enough to make any tunnelling term negligible, and thus, that states $\ket{L}$ and $\ket{R}$ can be treated as stationary states of the double well for the duration of the protocol. System $B$ is an ancillary qubit (AQ) system, which may have the same or a different physical origin as system $A$~\cite{gieseler2020,martinetz2020}. We stress that the argument that we will put forward is independent of the precise physical nature of system B. Finally, $C$ is a mediator particle of mass $m_c$ trapped in a harmonic potential characterised by an oscillation frequency ${\omega}$ in the X direction, and we assume that its motion in this dimension is uncoupled from its motion in orthogonal dimensions. A similar setup, albeit without system $B$, has been considered in Refs.~\cite{Taylor2020,Streltsov2021a}. Here, we assume that system $C$ interacts with both $A$ and $B$, while the interaction between the latter can be neglected. Under this 
assumption, the setup is well described by a Hamiltonian 
of the form
\begin{multline}
\label{eq:Ham}
    \hat{H} = \hbar \omega_0 \hat{\sigma}_b^z + \frac{1}{2m_c} \hat{P}^2 + \frac{1}{2} m_c \omega^2 \hat{X}^2 +\\
     \sum_{\alpha=L,R} \hat{V}_{a,\alpha}(\hat{X}) \ketbra{\alpha}{\alpha} + \sum_{\alpha=0,1} \hat{V}_{b,\alpha}(\hat{X}) \ketbra{\alpha}{\alpha},
\end{multline}
where $\hat{X}$ and $\hat{P}$ are, respectively, the position and momentum operators of the mediator, and $\hat{\sigma}_{b}^z$ is the Pauli $z$-operator acting on system $B$, with $\omega_0$ giving its bare energy splitting. The terms $\hat{V}_{i,\alpha}$, with $i=\{a,b\}$, represent the interaction energy between system $C$ and system $i$ when the latter is in state $\alpha$ and are assumed to be a function of the position of the mediator. We are interested in the case where $\hat{V}_{a,\alpha}$ is purely of gravitational origin, while $\hat{V}_{b,\alpha} \gg \hat{V}_{a, \alpha}$ and, although typically not of gravitational origin, its specific physical origin is not relevant for the argument. In order to avoid the interaction between $A$ and $C$ being dominated by Casimir forces, the distance between these masses needs to be sufficiently large---the precise value depending on their masses---typically exceeding significantly the splitting $d_0$ of the double-well potential~\cite{SuppMat}. Hence, we can expand the gravitational potential to second order in the separation distance around the value $d$ to find an interaction energy
\begin{multline}
\label{eq:Newton}
\hat{V}_{a,\pm}(\hat{X}) = - \frac{G m_a m_c}{\abs{d \mp \frac{d_0}{2} + \hat{X}} }  \\
\approx -\frac{G m_a m_c}{d} \left( 1 + \frac{d_0^2}{4d^2}  \pm \frac{d_0}{2d} - (1 \pm \frac{d_0}{d})\frac{\hat{X}}{d} +\frac{\hat{X}^2}{d^2} + ...\right),
\end{multline}
where $\hat{V}_{a,+}$ and $\hat{V}_{a,-}$ correspond, respectively, to $\hat{V}_{a,R}$ and $\hat{V}_{a,L}$, and $G = 6.67408 \cdot 10^{-11}$ m$^3$ kg$^{-1}$ s$^{-2}$ is the gravitational constant. The first two terms in the expansion introduce a global energy shift, the third gives an energy splitting of the TLTM, while the fourth term is responsible for a displacement of the oscillator equilibrium position and as well as for a linear interaction between mediator and the TLTM. Finally, the fifth term generates a shift in the oscillation frequency of the oscillator. Thus, putting everything together, Hamiltonian~(\ref{eq:Ham}) can be rewritten as
\begin{equation}
\label{eq:HamLinear}
\hat{H} = \hbar \omega_a \hat{\sigma}_a^z + \hbar \omega_b \hat{\sigma}_b^z + \hbar \tilde \omega \hat{a}^\dag \hat{a} + \hbar (g_a \hat{\sigma}_z^a + g_b \hat{\sigma}_z^b)(\hat{a} + \hat{a}^\dag)
\end{equation}
provided that the interaction energy between systems $C$ and $B$ admits a similar expansion, and that $\abs{\pm d_0/2 - \Delta_x} \ll d$, with  $\Delta_x$ denoting the maximum value of the position uncertainty of the oscillator during its evolution.  Here, $\hat{a}^\dag$ and $\hat{a}$ are ladder operators of the harmonic oscillator $C$ with modified frequency ${\tilde \omega}^2 = \omega^2 - \frac{2G m_a}{d^3} + \frac{2}{m_c}V^{(2)}_b$, where $V^{(2)}_b$ is the coefficient of the term quadratic in $\hat{X}$ in the interaction between $C$ and $B$. Furthermore, $\omega_a = Gm_a m_c d_0/(2\hbar d^2)$ and $g_a = - (Gm_a d_0/d^3)\sqrt{m_c/(2 \tilde \omega \hbar)}$, upon defining $\sigma_z^a = \ketbra{L}{L} - \ketbra{R}{R}$; and $\omega_b$ and $g_b$ have similar expressions in terms of the specific interaction between $B$ and $C$.

\begin{figure}[htbp] %  figure placement: here, top, bottom, or page
   \centering
   \includegraphics[width=\columnwidth]{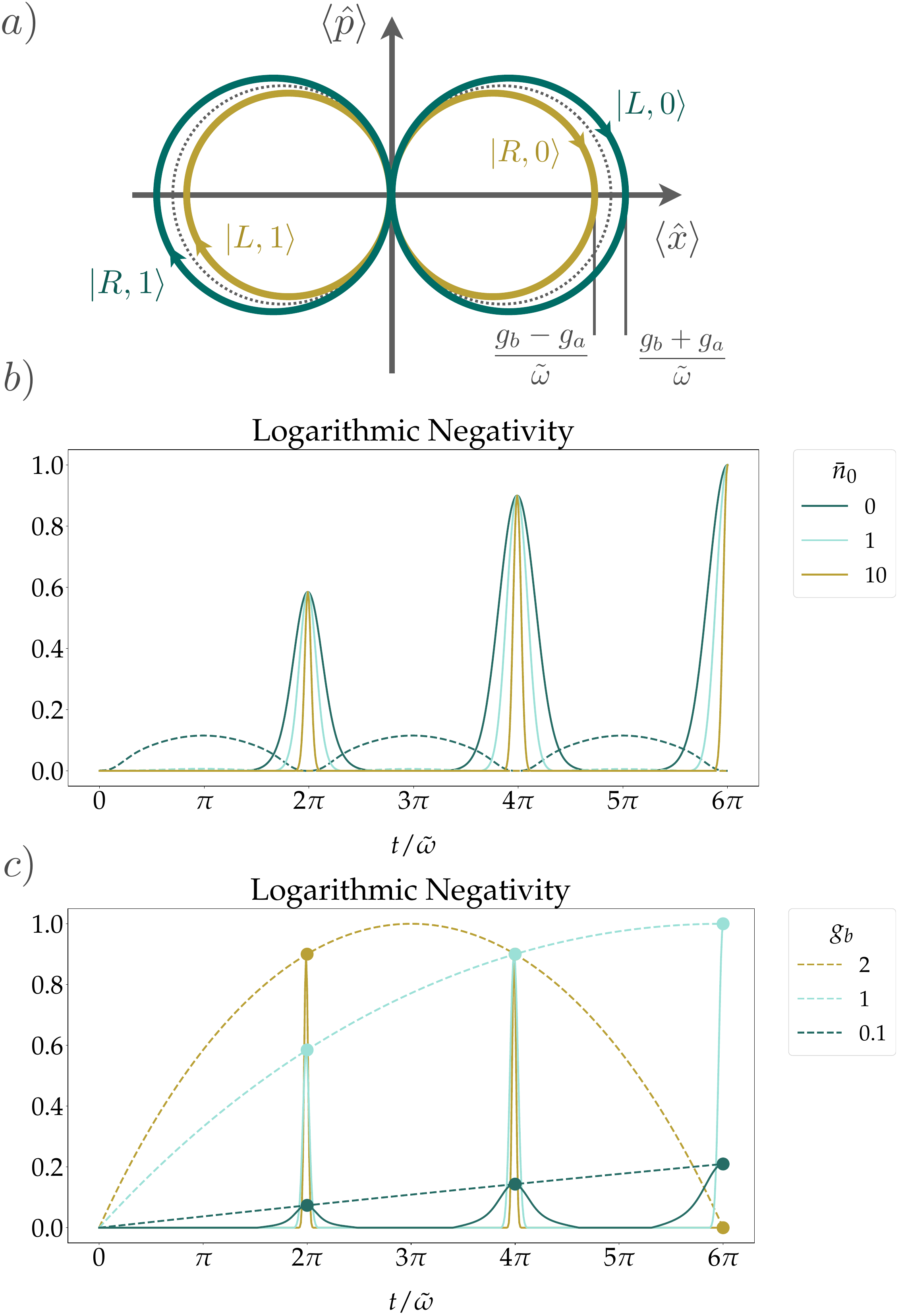} 
   \caption{{\bf System dynamics} In (a) we show the evolution in phase space of the four components of the mediator state correlated with each of the four states of the TLTM and the AQ. Here, $\{\expval{\hat{x}}, \expval{\hat{p}}\}$ are dimensionless position and momentum quadratures of the redefined oscillator, with shifted frequency and displaced equilibrium position. (b) shows the evolution of the entanglement, as quantified by the logarithmic negativity, between the TLTM and the AQ in continuous lines and between the TLTM and the mediator in dashed lines, for different temperatures of the mediator. Here, $g_a = 1/48 \tilde \omega$ and $g_b = \tilde \omega$. Continuous lines in (c) display the entanglement between the TLTM and the AQ for different values of the coupling $g_b$ expressed in units of $\tilde \omega$. For this simulation we initialize the mediator in a thermal state with mean phonon occupation $\expval{n}_0 = 10$, and set $g_a$ as in (b). Dots indicate the value of the entanglement at the decoupling times $t_n$. Dashed lines correspond to the evolution of entanglement between two generic qubits governed by $\hat{H} = \frac{2 g_a g_b}{\tilde \omega} \hat{\sigma}_z \hat{\sigma}_z$, which at times $t_n$ has a unitary-evolution operator equivalent to that of the full-system Hamiltonian, see Eq.~(\ref{eq:unitaryEvolution}).}
   \label{fig:dynamics}
\end{figure}

{\em Dynamics and Entanglement.---}The unitary-evolution operator associated to Hamiltonian~(\ref{eq:HamLinear}) can be conveniently expressed in the interaction picture as~\cite{SuppMat}
\begin{multline}
\label{eq:unitaryEvolution}
\hat{U}(t) = \exp{(g_a \hat{\sigma}_z^a + g_b \hat{\sigma}_z^b)(-\hat{a} \alpha_t + \hat{a}^\dag \alpha_t^*)} \\ \times \exp{-i\frac{2 g_a g_b}{\tilde \omega} \hat{\sigma}_z^a \hat{\sigma}_z^b (t - \frac{\sin{\tilde \omega t}}{\tilde \omega})},
\end{multline}
with $\alpha_t = \frac{e^{-i \tilde \omega t} - 1}{\tilde \omega}$. The first term generates a time-dependent displacement of the mediator in phase space conditional on the states of the TLTM and the AQ. The second term gives a second order interaction between the two lateral systems with an effective coupling $g_{\rm eff} = 2 g_a g_b / \tilde \omega$. Remarkably, at times $t_n = 2 \pi n/ \tilde \omega$ that are a natural period of the mediator frequency, $\alpha_{t_n} = 0$ and the first term vanishes, leaving an effective interaction between the TLTM and the AQ which is independent of the state of the oscillator, with $\hat{U}(t_n) = \exp{-i g_{\rm eff} \hat{\sigma}_z^a \hat{\sigma}_z^b t_n}$. Therefore, at these points in time the mediator is decorrelated from the rest of the system, while entanglement is retained between the TLTM and the AQ. Thus, provided that the TLTM and the AQ are initialized in suitable states, and that the gravitational interaction is able to mediate quantum correlations, entanglement will grow between the TLTM and AQ. This entanglement can then be detected by standard methods making local measurements on the 2-qubit system~\cite{Virmani07,HorodeckiKH09}. The principle that gives rise to the interaction is the same as that of the phase gates employed in trapped-ion platforms to entangle their internal degrees of freedom mediated by their collective motion~\cite{MolmerAS99, Solano99, Milburn00,SackettCM00}. Here, we use it as an amplification mechanism of the gravitational interaction. Notice, that the interaction strength between the TLTM and the AQ grows with the mass of the mediator as $\sqrt{m_c}$ and can be enhanced by a factor $g_b/\omega$ over the strength of the gravitational interaction $g_a$. The latter occurs because during the evolution the mediator will be displaced in phase space in opposite directions conditionally on the states of the TLTM and AQ, with this displacements reaching values of $(g_a + g_b)/\tilde \omega$, see Fig.~\ref{fig:dynamics}a. Thus, with increasing coupling of the AQ to the mediator, this grows into states with larger spatial delocalization, which in turn enhance the interaction between the TLTM and the mediator.

In practice, the tolerable delocalization of the mediator will be limited by the distance that preserves the linear approximation in the expansion of the gravitational potential that we did in Eq.~(\ref{eq:Newton}), that is $\Delta_x \approx \sqrt{\hbar/ (2 m \tilde \omega) } \sqrt{ \expval{\bar n}} \ll d$, where $\Delta_x$ and $\expval{\bar n}$ correspond, respectively, to the maximum values of the position uncertainty and the mean phonon occupation number of the mediator during the evolution. The time-dependent phonon occupation number $\expval{n}_t$ can be exactly calculated for an initial state with the TLTM and the AQ in even superpositions of the type $(\ket{L/0} + \ket{R/1})/\sqrt{2}$ and the mediator in a thermal state with mean phonon occupation number $\bar n_0$. It is given by
\begin{equation}
\expval{n}_t = \bar n_0 + 4 \frac{g_a^2 + g_b^2}{{\tilde \omega}^2} \sin^2{\frac{\tilde \omega t}{2}}.
\end{equation}
This sets a limit on the strength of the coupling of the AQ to the mediator
\begin{equation}
\label{eq:couplingBound}
g_b/\tilde \omega \approx \frac{1}{2} \sqrt{\frac{2 m_c \tilde \omega \Delta_x^2}{\hbar} - \bar n_0} \ll \frac{1}{2} \sqrt{\frac{2 m_c \tilde \omega d^2}{\hbar} - \bar n_0},
\end{equation}
where we have assumed $g_b \gg g_a$. 
We now consider the entanglement dynamics between the TLTM and the AQ, which we quantify in terms of the logarithmic negativity $LN = \max{ (0, \log _2{ {\lVert \rho_{ab}^{T_b}} \rVert}_1 )}$, with ${\norm{\cdot}}_1$ the trace norm and where $T_b$ represents the partial transpose with respect to subsystem $B$. For a closed system ruled by Hamiltonian~(\ref{eq:HamLinear}) an exact expression can be found at the times when the mediator is decoupled from the system: 
\begin{equation}
\label{eq:LogNeg}
 LN(t_n) = \max \{ 0, \log_2 [ 1 + \abs{\sin(\phi_m)} ] \},
 \end{equation} 
with 
\begin{equation}
\label{eq:EntMediated}
\phi_m = \frac{4 g_b g_a}{\tilde \omega} t_n \approx \frac{2 G m_c m_a}{\hbar d^3} \Delta_x d_0 t_n,
\end{equation}
where we have assumed for simplicity that $n_0 = 0$. This expression is upper bounded due to the constraint  $\Delta_x \ll d$. In Fig.~(\ref{fig:dynamics}b) we show the dynamics of entanglement between the different subsystems, for various temperatures of the mediator. We see that when the mediator starts in the GS the logarithmic negativity between the TLTM and the mediator oscillates with the period of the mediator frequency and vanishes completely at times $t_n$. At these times the mediator is decoupled while the entanglement between the TLTM and the AQ reaches a maximum. While the logarithmic negativity between the mediator and the TLTM decreases with increasing temperature, the entanglement between the TLTM and the AQ at the rephasing times $t_n$ remains unaffected. This is observed in the form of peaks centered at positions $t_n$, which get narrower with increasing temperature of the mediator. In Fig.~(\ref{fig:dynamics}c) we illustrate the enhancement of the entanglement between the TLTM and the AQ as the coupling of the AQ to the mediator is increased. We observe that with increasing coupling strength the peaks of entanglement between the TLTM and the AQ become higher and narrower. 

To understand the degree of amplification that such a setup can introduce, we compare it to the case without a mediator. We consider two gravitationally interacting TLTMs separated by a distance $D$, whose double-well potentials have a separation $d_0$. For the setup featuring a mediator we consider a TLTM with the same separation distance $d_0$, located a distance $d$ away from the mediator. The distance $d$ will in general be larger than $D$ by an amount given by the difference between the radii of the mediator, $R$, and the TLTM, $r$, that is $d = D + \Delta R - d_0/2 + \Delta_x/2$, with $\Delta R = R - r$. In this way we make sure that the distances between the surfaces of the gravitationally interacting bodies is the same in both setups, and thus avoid the appearence of Casimir-Polder forces between the mediator and the TLTM. We find that in the case of two directly interacting TLTMs the logarithmic negativity evolves as in Eq.~(\ref{eq:LogNeg}) with the argument of the sine given by \cite{SuppMat}
\begin{equation}
\phi_d = \frac{G m_a^2}{\hbar D}\frac{(d_0/D)^2 t}{1 - (d_0/D)^2}.
\end{equation}
Thus, the enhancement of the mediated setup over the setup with directly interacting masses can be expressed as the ratio
\begin{equation}
\label{eq:EnhancementRatio}
\phi_m/ \phi_d = 2 \frac{m_c \Delta_x}{m_a d_0} \frac{1}{(1 + \frac{\Delta R}{D})^3},
\end{equation}
where we have assumed $d_0/D \ll 1$. To quantitatively illustrate such an enhancement we examine the following example. Consider a particle of silica with radius $r = 70$~nm (recently, particles of this size have been placed in their motional GS~\cite{DelicRD2020, magrini2021}), in a double-well potential with $d_0 = 500$~nm. If we impose that the gravitational interaction energy has to exceed the Casimir interaction energy by a factor of $10$, we find that the minimum distance between their surfaces must be kept above $166\ \mu$m. This holds for all silica particles with radii below $166\ \mu$m \cite{SuppMat}). Thus,  we fix $D = 166\ \mu$m and consider a mediator with a radius that is $\alpha$ times lager than that of the TLTM, that is $R = \alpha r$. Assuming a frequency for the mediator of $\tilde \omega = 100$ Hz, and that both mediator and TLTM are silica particles, with mass density $\rho \approx 2400$ Kg/m$^3$, this gives an enhancement of 
\begin{equation}
    \phi_m/\phi_d \approx \frac{\sqrt{\alpha^3}}{[1 + (\alpha -1)4\cdot 10^{-4}]^3} 10^{-3} \frac{g_b}{\tilde \omega},
\end{equation}
Thus, we see that, for example, a mediator particle of radius $R = 7\ \mu$m, corresponding to $\alpha = 100$, would provide an enhancement on the order of $\phi_m/\phi_d \approx g_b/\tilde \omega \ll 10^8$, where the upper bound is imposed by the relation in Eq.~(\ref{eq:couplingBound}).

{\it Conclusion.---}The detection of gravitationally mediated entanglement would represent a remarkable experimental result with far-reaching consequences for our understanding of physics. Although this is an outstanding technological challenge that will require the quantum control of heavier and heavier systems, rapid developments and recent experimental breakthroughs in the field of optomechanics suggest that the consideration of this question is pertinent and timely. In this spirit, we propose an enhancement of the experimental design w.r.t.\@ existing proposals, which rely on the direct gravitational interaction between heavy test masses. In our design, we shift the large mass requirement to a mediator system, while keeping the test systems, where the entanglement is to be detected, at scales more friendly for their quantum control. While these smaller test systems would not show detectable amounts of entanglement were they to interact directly, in the mediated design, they show an effective interaction that grows with the mass of the mediator. Remarkably, the required degree of controllability on the heavy mediator mass is considerably lower than that of the test systems in the directly interacting case, such that the mediator can remain in a thermal state. This paves the way for experimental tests of the gravitational interaction between masses that are significantly larger than those that can be prepared in pure states. 

{\em Acknowledgments --} This work was supported by the ERC Synergy grant HyperQ (Grants No. 856432), the EU projects and AsteriQs (Grants No. 820394), the QuantERA project NanoSpin (Contract No. 13N14811), the BMBF project Q.Link.X (Contract No. 16KIS0875) and the DFG SFB 1279.

%%%%%%%%%% Merge with supplemental materials %%%%%%%%%%

\onecolumngrid
\clearpage

\begin{center}
\textbf{\large Supplemental Material:\\
Enhancing Gravitational Interaction between Quantum Systems by a Massive Mediator}
\begin{minipage}[c]{0.8\textwidth}
    \vspace{25pt}
We provide detailed analytical derivations of the time evolution of several of the magnitudes discussed in the main text. These include the unitary-evolution operator, the mean phonon number of the mediator, and the logarithmic negativity. We also describe our treatment of Casimir forces and propose a pulse sequence to suppress gravitational gradient noise. 
    \end{minipage}
\end{center}
%%%%%%%%%% Merge with supplemental materials %%%%%%%%%%
%%%%%%%%%% Prefix a "S" to all equations, figures, tables and reset the counter %%%%%%%%%%
\setcounter{equation}{0}
\setcounter{figure}{0}
\setcounter{table}{0}
\setcounter{page}{1}
\makeatletter
\renewcommand{\theequation}{S\arabic{equation}}
\renewcommand{\thefigure}{S\arabic{figure}}
\renewcommand{\bibnumfmt}[1]{[S#1]}
\renewcommand{\citenumfont}[1]{S#1}
%%%%%%%%%% Prefix a "S" to all equations, figures, tables and reset the counter %%%%%%%%%%

\section{Unitary evolution operator}

In the following, we show the derivation of the unitary evolution operator in Eq.~(\ref{eq:unitaryEvolution}) of the main text. We start from Hamiltonian~(\ref{eq:HamLinear}) and move to the interaction picture, to find
\begin{equation}
\hat{H}_I(t) = (g_a \hat{\sigma}_a^z + g_b \hat{\sigma}_b^z) (\hat{a} e^{- i \tilde \omega t}  + \hat{a}^\dag e^{i \tilde \omega t}),
\end{equation}
which has a commutator at different times given by
\begin{equation}
[ \hat{H}_I(t), \hat{H}_I(t') ] = (g_a \hat{\sigma}_a^z + g_b \hat{\sigma}_b^z)^2 (e^{i \tilde \omega (t' -t)} - e^{-i \tilde \omega(t' - t)}),
\end{equation}
and all higher-order commutators vanish. Thus, the evolution in the interaction picture is fully described by the first two orders in the Magnus expansion $\hat{U}_I = \exp{\hat{\Omega}^{(1)}(t) + \hat{\Omega}^{(2)}(t)}$, with
\begin{equation}
\hat{\Omega}^{(1)}(t) = - i \int_0^t dt' \hat{H}(t')  = -i (g_a \hat{\sigma}_a^z + g_b \hat{\sigma}_b^z) (\hat{a} \frac{e^{-i \tilde \omega t} - 1}{-i \tilde \omega} + \hat{a}^\dag \frac{e^{i \tilde \omega t} - 1}{i \tilde \omega})
\end{equation}
and
\begin{equation}
\hat{\Omega}^{(2)}(t) = -\frac{1}{2} \int_0^t dt' \int_0^{t'} dt'' [ \hat{H}(t'), \hat{H}(t'') ] = i (g_a \hat{\sigma}_a^z + g_b \hat{\sigma}_b^z)^2 (t/\tilde \omega - \frac{\sin \tilde \omega t}{{\tilde \omega}^2}).
\end{equation}

\section{Time evolution of the mean phonon number of the mediator} 
In the following we compute the time evolution of the mean phonon number $\expval{\hat{n}}_t=\expval{\hat{a}^\dag \hat{a}}_t$ of the mediator c.m.\@ as the system evolves under the unitary evolution operator in Eq.~(\ref{eq:unitaryEvolution}) of the main text. We observe that $\hat{n}$ commutes with the second term in Eq.~(\ref{eq:unitaryEvolution}), and thus we only need to care about the first term. For the Hilbert space of the mediator, this acts as a displacement operator $\hat{D}(\hat{\beta}) = \exp{\hat{\beta} \hat{a}^\dag - \hat{\beta}^* \hat{a}}$, with 
\begin{equation}
    \hat{\beta} = \alpha_t^*(g_a \hat{\sigma}_z^a + g_b \hat{\sigma}_z^b),
\end{equation}
and $\alpha_t = (e^{-i \tilde \omega t} - 1)/ \tilde \omega $. Thus, the time evolution of the number operator is given by
\begin{equation}
    \hat{n} (t) = \hat{D}(\hat{\beta}) \hat{n} \hat{D}^\dag(\hat{\beta}) = \hat{n} - (\hat{\beta}^* \hat{a} + \hat{\beta} \hat{a}^\dag) + \abs{\hat{\beta}}^2.
\end{equation}
Considering an initial state of the TLTM and the AQ in a separable state with each system in an even superposition $(\ket{0/L} + \ket{1/R})/\sqrt{2}$, we have that $\langle \hat{\sigma}_z^{a/b}\rangle =0$ and thus $\langle \hat{\beta} \rangle = 0$ and $\langle \lvert \hat{\beta}\rvert^2 \rangle = \abs{\alpha_t}^2(g_a^2 + g_b^2)$. Therefore, the time evolution of the mean phonon number is given by
\begin{equation}
    \expval{\hat{n}}_t = \expval{\hat{n}} + \frac{g_a^2 + g_b^2}{\tilde \omega^2} 4 \sin^2\left( \frac{\tilde \omega t}{2} \right).
\end{equation}

\section{Logarithmic Negativity for two interacting qubits}
Consider a pair of two-level systems, $1$ and $2$, interacting through a Hamiltonian of the form ($\hbar = 1$) $\hat{H} = \Omega \hat{\sigma}_z^{(1)} \hat{\sigma}_z^{(2)}$. The system is initialized in the state 
\begin{equation}
    {\ket{\psi}_0 = (\ket{0} + \ket{1})/\sqrt{2}\otimes (\ket{0} + \ket{1})/\sqrt{2}}.
\end{equation}    
Its state at time $t$ is then given by 
\begin{equation}
{\ket{\psi (t)} = 1/2 [ e^{-i \Omega t}(\ket{0,0} + \ket{1,1}) + e^{i \Omega t} (\ket{0,1} + \ket{1,0}) ]}.
\end{equation}
Its entanglement can be quantified by the logarithmic negativity
\begin{equation}
    LN = \max[0, \log_2{ \norm{\hat{\rho}^{T_2}}_1}],
\end{equation}
where $^{T_2}$ represents the partial transpose with respect to one of the two subsystems, and $\norm{\cdot}_1$ is the trace norm. Thus, we first need to find the partial transpose of $\hat{\rho} (t) = \ketbra{\psi(t)}{\psi(t)}$,
\begin{equation}
\label{eq:partialTranspose}
    \hat{\rho}^{T_2} =\frac{1}{4} \left( \begin{array}{cccc} 1 & e^{i 2 \Omega t} & e^{-i 2 \Omega t} & 1 \\  e^{-i 2 \Omega t} & 1 & 1 & e^{i 2 \Omega t} \\ e^{i 2 \Omega t} & 1 & 1 & e^{-i 2 \Omega t} \\ 1 & e^{-i 2 \Omega t} & e^{i 2 \Omega t} & 1  \end{array} \right),
\end{equation} 
and then compute its trace norm, which is defined as 
$\norm{\hat{\chi}}_1 = \Trace{\sqrt{\hat{\chi} \hat{\chi}^\dag}}$. To that end, we first look for the eigenvalues of $\hat{\rho}^{T_2} (\hat{\rho}^{T_2})^\dag$, which are given by
\begin{equation}
    \lambda_{1,2} = \frac{1}{4} \sin^2(2 \Omega t), \quad \lambda_3= \frac{1}{4} \sin^4(t \Omega), \quad \lambda_4 = \frac{1}{4} \cos^4(\Omega t),
\end{equation} 
and compute the trace norm as the sum of their positive square roots
\begin{equation}
    \norm{\hat{\rho}^{T_2}}_1 = 1 + \abs{\sin(2 \Omega t)}.
\end{equation}
Finally, the logarithmic negativity can be written as 
\begin{equation}
\label{eq:LogNeg2qubit}
    LN (t) = \max \{0, \log_2[1 + \abs{\sin(2 \Omega t)}] \}.
\end{equation}

\section{Two directly interacting TLTMs}
Here we compute the evolution of the logarithmic negativity for a pair of TLTMs, $A$ and $B$, with masses $m_a$ and $m_b$, that interact directly (without a mediator) via gravity. We assume that the centers of the double-well potentials are separated by a distance $D$ and that the distance between the wells in each system is $d_0$. We consider that the interaction Hamiltonian is diagonal in the basis $\{\ket{L}_A, \ket{R}_A \}\otimes \{\ket{L}_B, \ket{R}_B \}$, and its elements are determined by the Newtonian gravitational interaction energy of the system in each configuration
\begin{equation}
    H = \left( \begin{array}{cccc} 
    a & 0 & 0 & 0 \\
    0 & b & 0 & 0 \\
    0 & 0 & c & 0 \\
    0 & 0 & 0 & d
    \end{array} \right),
\end{equation}
with $a = d = - \frac{G m_a m_b}{D}$, $b = - \frac{G m_a m_b}{D + d_0}$ and $c=- \frac{G m_a m_b}{D - d_0}$. This Hamiltonian can be rewritten in the convenient form
\begin{equation}
    \hat{H} = (a + \frac{b+c}{2} \mathbf{\hat{I}}_4) + \frac{b-c}{4} \hat{\sigma}_z^a\otimes \mathbf{\hat{I}}_2  + \frac{c-b}{4} \mathbf{\hat{I}}_2 \otimes \hat{\sigma}_z^b + \frac{1}{2}(a - \frac{b+c}{2}) \hat{\sigma}_z^a \otimes \hat{\sigma}_z^b,
\end{equation}
with $\hat{\sigma}_z^{a/b} = \ketbra{L}{L}_{A/B} - \ketbra{R}{R}_{A/B}$. In this form, we can immediately see that for an initial state $\ket{+}_A\otimes \ket{+}_B$, with $\ket{+}_i = (\ket{L}_i + \ket{R}_i)/\sqrt{2}$, the time evolution of the logarithmic negativity is given by Eq.~(\ref{eq:LogNeg2qubit}), with \begin{equation}
    2 \Omega t = (a - \frac{b + c}{2})/ \hbar = - \frac{G m_a m_b}{\hbar D} \frac{t}{(D/d_0)^2 - 1}.
\end{equation}

\section{Casimir forces} 
In this section we present our treatment of the Casimir forces to compute the minimum separation distance that favors gravitational interaction over Casimir. We start with the expression for the Casimir potential 
between two spheres of radius $r_a$ and $r_b$ with a surface to surface distance $d_s$. Under the assumption that $r_a,r_b \ll d_s$ this is given by
\begin{equation}
    V_c(d_s) = \frac{23\hbar c r_a^3r_b^3}{4\pi d_s^7} \left(\frac{\epsilon_r-1}{\epsilon_r+2}\right)^2 
    \label{Casimir}
\end{equation}
where $\epsilon_r$ is the dielectric constant. For the same configuration the gravitational energy between the two spheres is given by 
\begin{equation}
    V_g (d_s) = -\frac{G m_a m_b}{d_s + r_a + r_b}
\end{equation}
Thus, the minimum distance that guarantees a ratio of the gravitational to the Casimir energy above $\beta < V_g(d_s)/V_c(d_s)$ is given by 
\begin{equation}
    d_s > \left[ \frac{207 \hbar c}{(4 \pi)^3 G \rho_a \rho_b} \left(\frac{\epsilon_r-1}{\epsilon_r+2}\right)^2 \beta \right]^{1/6},
\end{equation}
where $\rho_{a/b}$ are the mass densities of particles $A$ and $B$. Using the parameters for silica, i.e. $\epsilon_r\cong 4$ and $\rho_{a/b}\cong 2400$ kg m$^{-3}$, and imposing $\beta = 10$, we find $d_s > 166\ \mu$m. Thus, for particles with radii $r_{a/b} \ll 166\ \mu$m, this surface to surface distance must be kept. 

\section{Noise}
In the main text we have discussed the performance of our protocol in the absence of any sources of noise that could be acting both on the two-level systems and on the mediator. Naturally, in the presence of these noise sources, the finite coherence time of the setup will limit the performance of our protocol. Notice, that such a coherence time would be similar in an experiment that would not use the {\it big mass} as a mediator but look into its entanglement with one of the two-level systems, albeit in this case the mass should be prepared in a pure state. While a complete analysis of the noise would require a detailed discussion of specific implementation platforms and is out of the scope of this work, here, we will discuss the effect of noise that originates from the action of a conservative force acting on the system. This could be for example the action of gravitational or electromagnetic noise. For noise sources that are far as compared to the spatial extension of the wave function of any of the subsystems, their impact on each of the subsystems can be modelled as an interaction with a linearized field of an intensity that varies stochastically in time. Under these assumptions,  the system Hamiltonian is of the following structure
 \begin{equation}
 H = [\Omega_0^a + \eta_a(t)] \sigma_z^a + [\Omega_0^b + \eta_b(t)] \sigma_z^b + \tilde \omega a^\dag a + [ g_a \sigma_z^a + g_b \sigma_z^b + \eta_c(t) ] (a + a^\dag),
 \end{equation}
 where $\Omega_0^i$ and $g_i$ represent for the two-level system $i$ its energy splitting and  coupling to the mediator, respectively. Here, noise on each of the three systems is introduced in the form of the three independent stochastic functions of time $\eta_i (t)$, for $i = a,b,c$. The corresponding unitary-evolution operator can be computed exactly using the Magnus expansion and following the same recipe as that used for the case without noise in the main text. The resulting expression reads
 \begin{equation}
 \label{eq:unitaryop}
U(t_n) =  e^{-i [ \Omega_0^a t_n + \bar \eta_a(t_n) - g_a \tilde {\tilde \eta}_c(t_n) ] \sigma^a_z} e^{-i [ \Omega_0^b t_n + \bar \eta_b(t_n) - g_b \tilde {\tilde \eta}_c(t_n) ] \sigma^b_z} e^{-i [ \tilde \eta_c(t_n) a + \tilde \eta_c^*(t_n) a^\dag ]} e^{i \frac{g_a g_b}{\tilde \omega} t_n \sigma_z^a \sigma_z^b},
 \end{equation}
 where $\bar \eta_i (t) = \int_0^t dt' \eta_i (t') $, $\tilde \eta_i (t) = \int_0^t dt' \eta_i(t') e^{-i \tilde \omega t'}$ and $\tilde {\tilde \eta}_i (t) = \int_0^t dt' \int_0^{t'} dt'' [\eta_i(t') + \eta_i (t'')] \sin[\tilde \omega (t' - t'')]$. Remarkably, in the presence of linear noise, the disentanglement of the two-level systems and the mediator at the times $t_n$ is preserved. This is because while the presence of the gradient noise stochastically displaces the mediator in phase space, it does not distort the trajectories, and notably the relative distances, of the different components of the superposition. Therefore, these componentes, which are displaced in different directions depending on the state of the two-level systems, are all affected in the same way by this additional displacement, and thus still recombine into the same point, albeit now not at the origin of the phase space but at some stochastically displaced point, as captured by the displacement operator in the third exponential of Eq.~(\ref{eq:unitaryop}). If this would not be the case, and the mediator would not perfectly disentangle from the two-level systems due to the action of noise, this residual entanglement with the mediator would reduce the entanglement between the two-level systems in a way dependnet on the temperature of the mediator. Nevertheless, the noise acting on the position of the mediator effectively enters as dephasing noise in the two-level system $i$ through the stochastic phase $g_i \tilde {\tilde \eta}_c(t)$. Thus, while noise acting on the mediator will unavoidably limit the performance of the protocol, it is remarkable that this does not depend on its state, and therefore, that the temperature of the mediator does not represent and added source of noise nor amplify it. 
 
 \section{Dynamical Decoupling}
  Very much inspired by spin systems, we assume that the equivalent of a $\pi$-pulse operation is available for our TLTM, that is an operation that acts on the Hilbert space of the TLTM as a $\sigma_x$ operator. Then a dynamical decoupling sequence can be generated by applying $\pi$-pulses simultaneously on the TLTM and on the AQ. Such a sequence would be able to cancel all noise that fluctuates in time scales slower than the inter-pulse spacing. 

 We assume that such a flip operation, corresponding to the application of a $\sigma_x$ operation to the state, can be performed in a time scale much faster than the period of any other frequency in the Hamiltonian. In this case, the effect of such a pulse can be modeled by an instantaneous sign flip of the operator $\sigma^{a/b}_z \rightarrow  - \sigma^{a/b}_z$. Thus, in the presence of such pulse sequences on the TLTM and the AQ, Hamiltonian~(\ref{eq:HamLinear}) in the main text acquires the from
\begin{equation}
\label{eq:HamNoise}
    H = \hbar [\omega_a + \eta_a(t)] F_a(t) \sigma_a^z + \hbar [\omega_b + \eta_b(t)] F_b(t) \sigma_b^z  + \hbar \tilde \omega a^\dag a+ \hbar [g_a F_a(t)\sigma_z^a + g_b F_b(t) \sigma_z^b + \eta_c(t)] (a + a^\dag),
    \end{equation}
where $\eta_i(t)$ are stochastic functions of time representing gravitational gradient noise on each of the three systems. Here, $F_{a(b)} (t)$ represents the pulse sequences on system $A$ ($B$) and is a function that takes only values $1$ and $-1$, such that its sign flips with every pulse that is applied on the system. We are interested in the time evolution under Hamiltonian~(\ref{eq:HamNoise}) and the ability of functions $F_{a/b} (t)$ to attenuate the effects of noise. To that end, we follow the same recipe as for the evolution in the noiseless case. We first move into the interaction picture
\begin{equation}
    \hat{H}_{\rm int}^I(t) =  \hbar [g_a F_a(t)\hat{\sigma}_z^a + g_b F_b(t) \hat{\sigma}_z^b + \eta_c(t)] (\hat{a} e^{-i \tilde \omega t} + \hat{a}^\dag e^{i \tilde \omega t})
\end{equation}
and then find the two-time commutator of the time-dependent Hamiltonian
\begin{equation}
    [\hat{H}_{\rm int}^I(t_1),\hat{H}_{\rm int}^I(t_2)] = \hbar^2 g_a \chi_a (t_1, t_2) \sigma_z^a + g_b \chi_b(t_1,t_2) \sigma_z^b + g_a g_b \xi(t_1,t_2) \sigma_z^a \sigma_z^b + {\rm scalar},
\end{equation}
where
\begin{equation}
    \chi_{a/b} (t_1, t_2) = - 2 i [F_{a/b}(t_1) \eta_c(t_2)  + F_{a/b}(t_2) \eta_c(t_1)]\sin{[\tilde \omega (t_1 - t_2)]}
\end{equation}
and
\begin{equation}
    \xi (t_1,t_2) = -2 i [F_a(t_1) F_b(t_2) + F_a(t_2) F_b(t_1)]\sin{[\tilde \omega (t_1 - t_2)]}.
\end{equation}
Higher-order time commutators vanish and the evolution is again given by the first two orders of the Magnus expansion $\hat{U}_I = \exp \{\hat{\Omega}^{(1)}(t) + \hat{\Omega}^{(2)} (t) \}$, with
\begin{equation}
    \hat{\Omega}^{(1)}(t) = -i \{ \sum_{i=a,b} g_i \hat{\sigma}_z^i [\hat{a} \bar F_i (t) + \hat{a}^\dag \bar F_i^*(t)]  + [\hat{a} \bar \eta_c(t) + \hat{a}^\dag \bar \eta_c^* (t)] \}
\end{equation}
and
\begin{equation}
    \hat{\Omega}^{(2)} (t) = -\frac{1}{2} [ g_a \tilde \chi_a (t) \hat{\sigma}_z^a + g_b \tilde \chi_b(t) \hat{\sigma}_z^b + g_a g_b \tilde \xi(t) \hat{\sigma}_z^a \hat{\sigma}_z^b ]
\end{equation}
Here, the upper bar indicates the single integral $\bar h(t) = \int_0^t dt_1 h(t_1) e^{- \tilde \omega t_1}$ and the tilde indicates the double integral $\tilde h(t) = \int_0^t dt_1 \int_0^{t_1} dt_2 h(t_1,t_2)$. On the one hand, we are interested in designing pulse sequences $F_{a/b} (t)$ that preserve the decoupling of the mediator at times $t_n$, that is, we want sequences that satisfy $\bar F_{a/b} (t_n)=0$. On the other hand, we want that, at the decoupling times, these sequences cancel the effects of noise, $\tilde \chi_{a/b}(t_n) = 0$, while they retain the interaction between systems $A$ and $B$, that is $\tilde \xi(t_n) \neq 0$. Condition $\bar F_{a/b} (t_n)=0$  is satisfied provided that the sequence is symmetric under a displacement in time of half of a period of the mediator, that is $F_{a/b}(t) = F_{a/b} (t + \pi/\tilde \omega)$. On the other hand, condition $\tilde \chi_{a/b} (t_n) = 0$ is achieved provided that the noise function $\eta_c(t)$ remains constant over the duration of $2 \Delta t$, where $\Delta t$ is the interpulse spacing, that is to say, if the noise fluctuates slower than the periodicity of the pulses. Finally, the resonance condition $\tilde \xi(t_n) \neq 0$ by which the signal survives, is achieved provided that both of the two-level systems are flipped simultaneously.


\begin{thebibliography}{99}

\bibitem{Feynman57} D. Rickles and C. M. DeWitt, {\it The Role of Gravitation in Physics: Report from the 1957 Chapel Hill Conference (Max-Planck-Gesellschaft zur Förderung der Wissenschaften, Berlin, Germany, 2011)}, Chap. 23.

\bibitem{Schrodinger52} E. Schr\"odinger, {\em Are There Quantum Jumps}, \href{https://academic.oup.com/bjps/article/III/11/233/1456020}{British Journal for the Philosophy of
Science {\bf 3}, 233 (1952)}.

\bibitem{AspelmeyerKM14} M. Aspelmeyer, T. J. Kippenberg, and C. Marquardt, {\em Cavity Optomechanics}, \href{https://journals.aps.org/rmp/abstract/10.1103/RevModPhys.86.1391}{Rev. Mod. Phys. {\bf 86}, 1391 (2014)}.

\bibitem{TeufelDL2011} J. D. Teufel, T. Donner, D. Li, J. W. Harlow, M. S. Allman, K. Cicak, A. J. Sirois, J. D. Whittaker, K. W. Lehnert, and R. W. Simmonds, {\em Sideband cooling of micromechanical motion to the quantum ground state}, \href{https://doi.org/10.1038/nature10261}{Nature {\bf 475}, 359 (2011)}.

\bibitem{ChanMS2011} J. Chan, T. P. M. Alegre, A. H. Safavi-Naeini, J. T. Hill, A. Krause, S. Gr\"oblacher, M. Aspelmeyer, and O. Painter, {\em Laser cooling of a nanomechanical oscillator into its quantum ground state}, \href{https://doi.org/10.1038/nature10461}{Nature {\bf 478}, 89 (2011)}.

\bibitem{RiedingerWM2018} R. Riedinger, A. Wallucks, I. Marinkovi\'c, C. L\"oschnauer, M. Aspelmeyer, S. Hong, and S. Gr\"oblacher, {\em Remote quantum entanglement between two micromechanical oscillators}, \href{https://doi.org/10.1038/s41586-018-0036-z}{Nature {\bf 556}, 473 (2018)}.

\bibitem{OckeloenDP2018} C. F. Ockeloen-Korppi, E. Damsk\"agg, J.-M. Pirkkalainen, M. Asjad, A. A. Clerk, F. Massel, M. J. Woolley, and M. A. Sillanp\"a\"a, {\em Stabilized entanglement of massive mechanical oscillators}, \href{https://doi.org/10.1038/s41586-018-0038-x}{Nature {\bf 556}, 478 (2018)}.

\bibitem{MillenMP2020} J. Millen, T.~S. Monteiro, R. Pettit, and A.~N. Vamivakas, {\em Optomechanics with Levitated Particles}, \href {https://doi.org/10.1088/1361-6633/ab6100} { Rep. Prog. Phys.  {\bf 83},  026401 (2020)}.

\bibitem{DelicRD2020} U. Deli\'c, M. Reisenbauer, K. Dare, D. Grass, V. Vuleti\'c, N. Kiesel, and M. Aspelmeyer, {\em Cooling of a levitated nanoparticle to the motional quantum ground state}, \href{https://doi.org/10.1126/science.aba3993}{Science {\bf 367}, 892 (2020)}.

\bibitem{CatanoSE2020} S. B. Cata\~no-Lopez, J. G. Santiago-Condori, K. Edamatsu, and N. Matsumoto, {\em High-Q Milligram-Scale Monolithic Pendulum for Quantum-Limited Gravity Measurements}, \href{https://doi.org/10.1103/PhysRevLett.124.221102}{Phy. Rev. Lett. {\bf 124}, 221102 (2020)}.

\bibitem{Whittle21} C. Whittle et al., {\it Approaching the motional ground state of a 10 kg object}, \href{https://doi.org/10.1126/science.abh2634}{Science {\bf 372}, 1333 (2201)}.

\bibitem{KotlerJT+2021} S. Kotler, A. Peterson, E. Shojaee, F. Lecocq, K. Cicak, A. Kwiatkowski, S. Geller, S. Glancy, E. Knill, R. W. Simmonds, J. Aumentado, and J. D. Teufel, {\it Direct observation of deterministic macroscopic entanglement}, \href{https://doi.org/10.1126/science.abf2998}{Science {\bf 372}, 622 (2021)}.

\bibitem{MercierMS+2021} L. Mercier de L\' epinay, C. F. Ockeloen-Korppi, M. J. Woolley, M. A. Sillanp\" a\" a, {\it Quantum mechanics–free subsystem with mechanical oscillators}, \href{https://doi.org/10.1126/science.abf5389}{Science {\bf 372}, 625 (2021)}.

\bibitem{SchmoleDH+16} J. Schm{\"o}le, M. Dragosits, H. Hepach, and M. Aspelmeyer, {\em A micromechanical proof-of-principle
experiment for measuring the gravitational force of milligram masses}, \href{https://iopscience.iop.org/article/10.1088/0264-9381/33/12/125031}{Class. Quantum Grav. {\bf 33}, 125031 (2016)}.

\bibitem{KafriT13} D. Kafri and J. M. Taylor, {\it A noise inequality for classical forces}, \href{https://arxiv.org/pdf/1311.4558.pdf}{arXiv:1311.4558}.

\bibitem{KrisnandaTP2017} T. Krisnanda, M. Zuppardo, M. Paternostro, and T. Paterek, {\it Revealing Nonclassicality of Inaccessible Objects}, \href{https://doi.org/10.1103/PhysRevLett.119.120402}{Phys. Rev. Lett. {\bf 119}, 120402 (2017)}.

\bibitem{Milburn2017} S. Bose, A. Mazumdar, G. W. Morley, H. Ulbricht, M. Toro\v{s}, M. Paternostro, A. A. Geraci, 
P. F. Barker, M. S. Kim, and G. Milburn, {\it Spin Entanglement Witness for Quantum Gravity}, \href{https://doi.org/10.1103/PhysRevLett.119.240401}{Phys. Rev. Lett. {\bf 119}, 240401 (2017)}.

\bibitem{MarlettoVV17} C. Marletto and V. Vedral, {\it Gravitationally Induced Entanglement between Two Massive Particles is Sufficient Evidence of Quantum Effects in Gravity}, \href{https://doi.org/10.1103/PhysRevLett.119.240402}{Phys. Rev. Lett. {\bf 119}, 240402 (2017)}.

\bibitem{MiaoMY+20} H. Miao, D. Martynov, H. Yang, and A. Datta, {\em Quantum correlation of light mediated by gravity}, \href{https://journals.aps.org/pra/abstract/10.1103/PhysRevA.101.063804}{Phys. Rev. A {\bf 101}, 063804 (2020)}.

\bibitem{KrisnandaTP+20} T. Krisnanda, G. Y. Tham, M. Paternostro, and T. Paterek, {\em Observable quantum entanglement due to gravity}, \href{https://www.nature.com/articles/s41534-020-0243-y}{npj Quant. Inf. {\bf 6}, 12 (2020)}.

\bibitem{PedernalesMP20b} J. S. Pedernales, F. Cosco, and M. B. Plenio, {\it Decoherence-Free Rotational Degrees of Freedom for Quantum Applications}, \href{https://doi.org/10.1103/PhysRevLett.125.090501}{Phys. Rev. Lett. {\bf 125}, 090501 (2020)}.

\bibitem{PedernalesMP20} J. S. Pedernales, G. W. Morley, and M. B. Plenio, {\em Motional Dynamical Decoupling for Matter-Wave
Interferometry.} \href{https://journals.aps.org/prl/abstract/10.1103/PhysRevLett.125.023602}{Phys. Rev. Lett. {\bf 125}, 023602 (2020)}; J. S. Pedernales, G. W. Morley, and M. B. Plenio, \href{https://arxiv.org/abs/1906.00835}{arXiv:1906.00835}.

\bibitem{CoscoMP20} F. Cosco, J. S. Pedernales, and M. B. Plenio, {\it Enhanced force sensitivity and entanglement in periodically driven optomechanics}, \href{https://doi.org/10.1103/PhysRevA.103.L061501}{Phys. Rev. A {\bf 103}, 061501 (2021)}.

\bibitem{Weiss2020}  T. Weiss, M. Roda-Llordes, E. Torrontegui, M. Aspelmeyer, and O. Romero-Isart, {\em Large Quantum Delocalization of a Levitated Nanoparticle Using Optimal Control: Applications for Force Sensing and Entangling via Weak Forces}, \href{https://doi.org/10.1103/PhysRevLett.127.023601}{Phys. Rev. Lett. {\bf 127}, 023601 (2021)}.

\bibitem{LindnerP2005} N. Lindner and A. Peres, {\em Testing quantum superpositions of the gravitational field with Bose-Einstein condensates}, \href{https://doi.org/10.1103/PhysRevA.71.024101}{Phys. Rev. A {\bf 71}, 024101 (2005)}.

\bibitem{BahramiBM+15} M. Bahrami, A. Bassi, S. McMillen, M. Paternostro, H. Ulbricht, {\em Is Gravity Quantum?} 
\href{https://arxiv.org/abs/1507.05733}{arXiv:1507.05733}. 

\bibitem{CarlessoPU+17} M. Carlesso, M. Paternostro, H. Ulbricht, and A. Bassi, {\em When Cavendish meets Feynman: A 
quantum torsion balance for testing the quantumness of gravity.} \href{https://arxiv.org/abs/1710.08695}{arXiv:1710.08695}.

\bibitem{CarlessoHU19} M. Carlesso, A. Bassi, M. Paternostro, and H. Ulbricht, {\em Testing the gravitational field generated by a quantum superposition}, \href{https://iopscience.iop.org/article/10.1088/1367-2630/ab41c1}{New J. Phys. {\bf 21}, 093052 (2019)}

\bibitem{Haine2021} S. A. Haine, {\it Searching for signatures of quantum gravity in quantum gases}, \href{https://doi.org/10.1088/1367-2630/abd97d}{New J. Phys. {\bf 23}, 033020 (2021)}.

\bibitem{Plenio05} M. B. Plenio, {\em The logarithmic negativity: A full entanglement monotone that is not convex}, \href{https://journals.aps.org/prl/abstract/10.1103/PhysRevLett.95.090503}{Phys. Rev. Lett. {\bf 95}, 090503 (2005)}.

\bibitem{StreltsovMP21} K. Streltsov, J. S. Pedernales, and M. B. Plenio, {\it Ground-State Cooling of Levitated Magnets in Low-Frequency Traps}, \href{https://doi.org/10.1103/PhysRevLett.126.193602}{Phys. Rev. Lett. {\bf 126}, 193602 (2021)}.

\bibitem{vandenKamp2020} T.~W.~van de Kamp, R.~J.~Marshman, S.~Bose and A.~Mazumdar, {\it Quantum Gravity Witness via Entanglement of Masses: Casimir Screening}, \href{https://journals.aps.org/pra/abstract/10.1103/PhysRevA.102.062807}{Phys. Rev. A \textbf{102}, 062807 (2020)}.

\bibitem{Torov2020} M.~Toro\v{s}, T.~W.~Van De Kamp, R.~J.~Marshman, M.~S.~Kim, A.~Mazumdar, and S.~Bose, {\it Relative Acceleration Noise Mitigation for Nanocrystal Matter-wave Interferometry: Application to Entangling Masses via Quantum Gravity}, \href{https://doi.org/10.1103/PhysRevResearch.3.023178}{Phys. Rev. Research {\bf 3}, 023178 (2021)}.

\bibitem{RijavecCM21} S. Rijavec, M. Carlesso, A. Bassi, V. Vedral, and C. Marletto, {\it Decoherence effects in non-classicality tests of gravity}, \href{https://doi.org/10.1088/1367-2630/abf3eb}{New J. Physics {\bf 23}, 043040 (2021)}.

\bibitem{gieseler2020} J. Gieseler, A. Kabcenell, E. Rosenfeld, J. D. Schaefer, A. Safira, M. J. A. Schuetz, C. Gonzalez-Ballestero, C. C. Rusconi, O. Romero-Isart, and M. D. Lukin, {\it Single-Spin Magnetomechanics with Levitated Micromagnets}, \href{https://doi.org/10.1103/PhysRevLett.124.163604}{Phys. Rev. Lett. {\bf 124}, 163604 (2020)}.

\bibitem{martinetz2020} L. Martinetz, K. Hornberger, J. Millen, M. S. Kim, and B. A. Stickler, {\it Quantum Electromechanics with Levitated Nanoparticles}, \href{https://doi.org/10.1038/s41534-020-00333-7}{Npj Quant. Inf. {\bf 6}, 101 (2020)}.

\bibitem{Taylor2020}  D. Carney, H. M{\"u}ller, and J. M. Taylor, {\em Testing quantum gravity with interactive information sensing}, \href{https://doi.org/10.1103/PRXQuantum.2.030330}{PRX Quantum {\bf 2}, 030330 (2021)}.

\bibitem{Streltsov2021a} K. Streltsov, J. S. Pedernales, and M. B. Plenio, {\it On the Significance of Interferometric Revivals for the Fundamental Description of Gravity}, \href{https://doi.org/10.3390/universe8020058}{Universe {\bf 8}, 58 (2022)}.

\bibitem{SuppMat} See Supplemental Material for a detailed derivation of the magnitudes discussed in the main text.

\bibitem{Virmani07} M. B. Plenio and S. Virmani, {\it An introduction to entanglement measures}, \href{http://www.rintonpress.com/journals/doi/QIC7.1-2-1.html}{Quant. Inf. Comp. {\bf 7}, 1 (2007)}. 

\bibitem{HorodeckiKH09} R. Horodecki, P. Horodecki, M. Horodecki, and K. Horodecki, {\it Quantum entanglement}, \href{https://doi.org/10.1103/RevModPhys.81.865}{Rev. Mod. Phys. {\bf 81}, 865 (2009)}.

\bibitem{MolmerAS99} K. M\o lmer and Anders S\o rensen, {\it Multiparticle Entanglement of Hot Trapped Ions}, \href{https://doi.org/10.1103/PhysRevLett.82.1835}{Phys. Rev. Lett. {\bf 82}, 1835 (1999)}.

\bibitem{Solano99} E. Solano, R. L. de Matos Filho, and N. Zagury, {\it Deterministic Bell states and measurement of the motional state of two trapped ions}, \href{https://doi.org/10.1103/PhysRevA.59.R2539}{Phys. Rev. A {\bf 59}, R2539 (1999)}.

\bibitem{Milburn00} G. J. Milburn, S. Schneider and D. F. V. James, {\it Ion Trap Quantum Computing with Warm Ions}, \href{https://onlinelibrary.wiley.com/doi/10.1002/1521-3978(200009)48:9/11<801::AID-PROP801>3.0.CO;2-1}{Fortschr. Phys. {\bf 48}, 801 (2000)}.

\bibitem{SackettCM00} C. A. Sackett, D. Kielpinski, B. E. King, C. Langer, V. Meyer, C. J. Myatt, M. Rowe, Q. A. Turchette, W. M. Itano, D. J. Wineland, and C. Monroe, {\it  Experimental entanglement of four particles}, \href{https://doi.org/10.1038/35005011}{Nature {\bf 404}, 256 (2000)}.

\bibitem{magrini2021} L. Magrini, P. Rosenzweig, C. Bach, A. Deutschmann-Olek, S. G. Hofer, S. Hong, N. Kiesel, A. Kugi, and M. Aspelmeyer, {\em Real-time optimal quantum control of mechanical motion at room temperature}, \href{https://doi.org/10.1038/s41586-021-03602-3}{Nature {\bf 595}, 373 (2021)}.


\end{thebibliography}
\end{document}